\documentclass[aps,pra,twocolumn]{revtex4}
\newcommand {\be}{\begin{equation}}
\newcommand {\ee}{\end{equation}}
\newcommand {\bey}{\begin{eqnarray}}
\newcommand {\eey}{\end{eqnarray}}
\usepackage{amssymb}
\usepackage{amsmath}
\usepackage{amsfonts}
\usepackage{mathbbol}
\usepackage{hyperref}
\usepackage{epsfig}
\usepackage{bbold}
\usepackage{dsfont}
\newtheorem{theorem}{Theorem}
\newtheorem{lemma}{Lemma}
\newtheorem{property}{Property}
\newtheorem{definition}{Definition}

\newcommand{\qed}{\nobreak \ifvmode \relax \else
      \ifdim\lastskip<1.5em \hskip-\lastskip
      \hskip1.5em plus0em minus0.5em \fi \nobreak
      \vrule height0.75em width0.5em depth0.25em\fi}

\begin{document}

\title{Generalized Gleason theorem and finite amount of information for the context}

\author{A. Montina}
\author{S. Wolf}
\affiliation{Facolt\`a di Informatica, Universit\`a della Svizzera Italiana, Via G. Buffi 13, 6900 Lugano,
Switzerland}

\date{\today}

\begin{abstract}
Quantum processes cannot be reduced, in a nontrivial way, to classical processes
without specifying the context in the description of a measurement procedure. 
This requirement is implied by the Kochen-Specker theorem in the outcome-deterministic
case and, more generally, by the Gleason theorem. The latter establishes that there is
only one non-contextual classical model compatible with quantum theory, the one that trivially identifies
the quantum state with the classical state. However, this model requires a breaking of
the unitary evolution to account for macroscopic realism. Thus, a causal classical model
compatible with the unitary evolution of the quantum state is necessarily contextual
at some extent. Inspired by well-known results in quantum communication complexity, we 
consider a particular class of hidden variable theories by assuming that the amount of 
information about the measurement context is finite. Aiming at establishing some
general features of these theories, we first present a generalized version of the Gleason 
theorem and provide a simple proof 
of it. Assuming that Gleason's hypotheses hold only locally for `small' changes of the 
measurement procedure, we obtain almost the same conclusion of the original theorem about 
the functional form of the probability measure. An additional constant and a relaxed property 
of the `density operator' are the only two differences from the original result. By this
generalization of the Gleason theorem and the assumption of finite information for the
context, we prove that the probabilities over three or more outcomes of a 
projective measurement
must be linear functions of the projectors associated with the outcomes, given the 
information on the context. 
\end{abstract}
\maketitle

\section{introduction}

In the formalism of quantum theory, the possible outcomes of a von Neumann measurement are 
labeled by projectors. This description provides an operationally exhaustive summary of the 
whole measurement procedure and contains the complete information that is relevant for 
distinguishing two events that occur with different probability for some preparation
procedure. Furthermore, this labeling 
is also minimal, that is, it does not distinguish events that always occur with the same 
probability. Given a measurement, the outcomes are identified by a set of $M$ commuting 
projectors, say $\{\hat E_1,\hat E_2,\cdots,\hat E_M\}$, with $\sum_{n=1}^M \hat E_n=\mathbb{1}$. 
The probability of 
outcome $\hat E_k$, say $\mu(\hat E_k)$, is given by the Born rule
\be
\mu(\hat E_k)=\text{Tr}(\hat E_k\hat\rho),
\ee
where $\hat\rho$ is the density operator, which gives the statistically relevant
information about the preparation procedure. The additional information about
the measurement procedure that is irrelevant for the computation of $\mu(\hat E_k)$
is referred as {\it context}. For example, the projectors
$\hat E_2,\hat E_3,\dots,\hat E_M$ are part of the context for the outcome
$\hat E_1$, as they are not relevant for computing $\mu(\hat E_1)$.

Although this formalism is operationally exhaustive and minimal,
it does not provide a unified
description of observed and observing systems. Indeed, two different languages 
are used for the experimental apparatus and the quantum system under observation. 
On the one hand the experimental 
apparatus is described by a purely classical language that specifies for example 
the position and orientation of beam splitters, mirrors or crystals in a quantum 
optics experiment. On the other hand the quantum system is indirectly described 
by the operations performed on the experimental apparatus. The quantum state is 
not meant as a classical object, such as a field, but it is a mere container of 
information about the preparation procedure. This formalism is completely silent 
on the actual state of affairs of each single quantum system. Is it possible to 
have a unified description that puts experimental apparatus and quantum system 
on the same footing? Various no-go theorems show that this embedding of quantum
processes in the classical framework is not possible without apparently unphysical
consequences, such as non-locality~\cite{bell1,bell2} and, more generally, 
contextuality~\cite{kochen}. The latter refers to a dependence of the
outcomes on the details of the measurement implementation that are irrelevant 
for the computation of the quantum probabilities $\mu(\hat E_k)$.
Apart from foundational motivations concerning the interpretation of the quantum formalism,
a classical embedding of quantum theory is important in the context of quantum information 
theory, since no-go theorems concerning this embedding make the gap between quantum and 
classical information more definite. For example, Bell theorem led to the discovery of 
quantum cryptographic protocols exploiting non-locality as a resource~\cite{barrett}.
In quantum communication complexity, classical simulations are
relevant for setting a limit on the advantages offered by quantum channels.

Hidden variable (HV) theories~\cite{mermin}, also known as ontological 
theories, are one possible classical reinterpretation of quantum
processes. In a HV theory, there is no 
dichotomy between classical and microscopic quantum world. Any system is always 
supposed to be in some well-defined classical state, say $x$, which is an element 
of a classical space, $X$. According to the present terminology~\cite{spekkens}, 
we will refer to the 
classical state and the classical space as {\it ontic state} and {\it ontological 
space}, respectively. Given a preparation procedure, the system is set in an
ontic state according to some probability distribution that depends on the 
preparation procedure. When a measurement is performed, the probability of an 
outcome is conditioned by the ontic state. If the outcome is completely determined, 
then the ontological theory is said to be {\it outcome-deterministic}. 

The de Broglie-Bohm (dBB) theory is a particular example of outcome-deterministic HV 
theory. In this case, the quantum state assumes the role of an actual physical field 
that pilots the dynamics of the particles. Thus, the ontic state $x$ is identified 
with the wave-function and the positions of the particles.  Another example is given 
by the Beltrametti-Bugajski (BB) model. Differently from dBB theory, the ontic state
is identified with only the quantum state, which is not supplemented by any additional
variable. Furthermore, the BB model is 
not outcome-deterministic. In more general HV theories, the ontic state does not 
necessarily contain the full information about the quantum state, which is instead 
encoded into the statistical behavior of many identically prepared realizations.
We call an ontological theory {\it trivial} if, for any measurement, the outcome 
probabilities, given an ontic state $x$, are equal to the quantum probabilities, 
given some quantum state $|\psi\rangle$. In this case, $x$ can be identified with 
$|\psi\rangle$. According to this definition, the BB model is a trivial ontological 
theory. Any trivial HV theory is essentially equivalent to the BB model.
The dBB theory, being deterministic, is a counterexample of nontrivial HV theory.

In their seminal article \cite{kochen}, Kochen and Specker showed that any 
outcome-deterministic ontological theory is measurement-contextual. In 
other words, the minimal labeling of an event with a projector is not
sufficient to describe consistently a measurement procedure. In Ref. 
\cite{spekkens}, it was pointed out that the outcome determinism is 
a necessary condition for inferring the measurement contextuality.
Indeed the BB model is an example of measurement-noncontextual ontological
theory, which is not outcome-deterministic. 
In fact the BB model is the only noncontextual ontological theory~\cite{chen}. 
Equivalently, quantum mechanics is essentially the only theory that employs 
the minimal labeling in the description of the events. In Ref.~\cite{chen}, we 
argued that the BB model, being a trivial HV theory, is not sufficient for 
introducing realism in the quantum phenomena, unless the unitarity 
of the evolution is broken. Thus, we concluded that the measurement contextuality 
should be introduced to some extent. What is the minimal amount of required information 
about the  context?  It is a well-known 
result of quantum information that a finite amount of classical communication is 
sufficient to reproduce classically the quantum correlations in an 
Einstein-Podolsky-Rosen (EPR) experiment~\cite{toner}. This kind of correlations 
provides a particular example of measurement contextuality~\cite{mermin}.
Inspired by this result of quantum information, we consider a particular class
of HV theories by assuming that the amount of relevant information about the 
measurement context is always finite. Using this hypothesis and a generalized version 
of the Gleason theorem, we prove that the probability of an event must be a 
linear function of the projector associated with the event, given the 
information on the context. 
This result provides an example illustrating the relevance of the generalized 
Gleason theorem. As remarked in the conclusion, this theorem can turn to be useful 
for solving some general questions in quantum information.

The paper is organized as follows. In Sec. \ref{sec_context} we introduce the 
hypothesis that the amount of relevant information about the context in a HV 
theory is finite. By this hypothesis, we show that the ontological theory is 
somehow noncontextual for `small' changes of the measurement procedure. In Sec. 
\ref{sec_gen_gleason}, we prove the generalized Gleason theorem. Assuming that 
Gleason's hypotheses hold only locally for `small' changes of the measurement 
procedure, we obtain almost the same conclusion of the original theorem about 
the functional form of the probability measure. An additional constant and a 
relaxed property of the `density operator' are the two only differences from 
the original result. The proof is much simpler than Gleason's proof. Furthermore 
the Gleason theorem can be derived as a corollary from this generalization.
In Sec. \ref{sec_prob_meas}, we derive the general form of the probability
of an event in a HV theory by using the hypothesis in Sec. \ref{sec_context}
and the generalized Gleason theorem. 
Finally, the conclusion and the perspectives are drawn.

\section{Finite amount of information about the context}
\label{sec_context}

Let us first introduce the framework of an ontological theory. 
For convenience, in the following we will associate projective measurements with
ordered $M$-tuples of projectors. A 
quantum system is described by an ontic state, $x$, which is an element of 
an ontological space, $X$. When the quantum system is prepared in a quantum 
state $|\psi\rangle$, its ontic state is set according to a probability 
distribution $\rho(x|\psi,\eta)$ which depends on $|\psi\rangle$ and,
possibly, an additional parameter, $\eta$, representing the preparation context.
Thus, we have the mapping
\be
(|\psi\rangle,\eta)\rightarrow \rho(x|\psi,\eta),
\ee
where $(|\psi\rangle,\eta)$ represents the preparation procedure. When a 
measurement ${\cal M}=(\hat E_1,\hat E_2,\cdots,\hat E_M)$ is performed, 
the probability of an outcome $\hat E_k$, is conditioned by the value of $x$. 
Differently from the quantum formalism, in general the probability also depends on 
the whole set $\cal M$, not just $\hat E_k$. The other projectors give the 
measurement context for the event $\hat E_k$. This dependence is not the only 
possible kind of contextuality. We denote by $\tau$ the additional context. 
Thus, a measurement procedure is specified by the pair $({\cal M},\tau)$, 
which is associated with a conditional probability of having outcome $\hat E_k$
given $x$, that is,
\be\label{cond_prob}
({\cal M},\tau)\rightarrow P(\hat E_k|x,{\cal M},\tau).
\ee
The set $\cal M$ is {\it complete}, that is, the sum of the projectors in $\cal M$
is the identity operator,
\be
\sum_k \hat E_k=\hat {\mathbb{1}}.
\ee
The probability distribution $P$ satisfies the normalization equation
\be
\sum_{k=1}^M P(\hat E_k|x,{\cal M},\tau)=1.
\ee
The ontological model reproduces a process of state preparation and subsequent
measurement if the equality
\be
\int dx P(\hat E_k|x,{\cal M},\tau)\rho(x|\psi,\eta)=\langle\psi|\hat E_k|\psi\rangle
\ee
is satisfied, where the integral is defined according to some measure on $X$.
In quantum communication complexity, this equation describes the simulation of 
a noiseless quantum channel with subsequent projective measurement.

Hereafter, we consider the class of ontological theories for which the amount of 
relevant information about the measurement context is finite. Thus,
we assume that the conditional probability in Eq. (\ref{cond_prob}) takes 
the form
\be\label{cond_prob2}
P(\hat E_k|x,{\cal M},\tau)=\sum_n\mu(\hat E_k|x,n)
P_c(n|x,{\cal M},\tau),
\ee
If the worst-case amount of information is finite, then the sum in Eq.~(\ref{cond_prob2})
is over a finite number of elements. More generally, we only assume that the index
$n$ is discrete and the
summation $\sum_n P_c(n|x,{\cal M},\tau)$ converges to $1$.
Note that the probability distribution $P_c$ explicitly depends on the ontic state, 
that is, the information about the context generally depends on the value $x$ in 
each single realization. Indeed, in the EPR scenario, this dependence is necessary 
if the context is summarized by a finite amount of information \cite{massar}. Also 
note that there is a redundancy in the definition of $\mu(\hat E_k|x,n)$, since
the index $n$ can contain some information about $\hat E_k$. This implies that there 
could be a conflict between the value of $n$ and $\hat E_k$. For example, if the 
probability $P_c(n'|x,{\cal M},\tau)$ is equal to zero for some $n'$
and for every $({\cal M},\tau)$ such that $E_k$ is equal to some $E_k'$, then
$\mu(\hat E_k'|x,n')$ is left indeterminate. Indeed, denoting by 
$P(\hat E_k,n|x,{\cal M},\tau)$ the joint probability of $\hat E_k$ and $n$, we have that
$$
\mu(\hat E_k'|x,n')=\frac{P(\hat E_k',n'|x,{\cal M},\tau)}{P_c(n'|x,{\cal M},\tau)}
\text{  for  }  \hat E_k=\hat E_k'
$$
and both the numerators and denominators are zero.
This point is important for correctly deriving the properties of the 
conditional probability $\mu(\hat E_k|x,n)$. In particular, the normalization condition 
$\sum_k \mu(\hat E_k|x,n)=1$ is required only if $n$ is consistent with $\hat E_k$ 
for every $k$, that is, if $P_c(n|x,{\cal M},\tau)\ne0$ for some $\tau$. Similarly,
the non-negativity condition $\mu(\hat E_k|x,n)\ge 0$ holds if $\hat E_k$ and $n$ are
consistent.

To state concisely the normalization condition for $\mu(\hat E_k|x,n)$, let us
introduce some set definition. We denote by $\Omega$ the set containing all the elements 
$\cal M$. This set, endowed with a 
Riemannian metric, is a Riemannian manifold. It has disjoint subsets and each subset 
contains elements whose projectors $\hat E_k$ have fixed rank. 
\begin{definition}
$\Omega_n(x)$ is the the largest subset of $\Omega$ such that, for every 
${\cal M}\in \Omega_n(x)$, $P_c(n|x,{\cal M},\tau)\ne0$ for some $\tau$.
\end{definition}
In other words, the set $\Omega_n(x)$ contains all the measurements that
are consistent with the context index $n$.
The sets $\Omega_n(x)$ cover the set $\Omega$, that is,
\be
\cup_n\Omega_n(x)=\Omega.
\ee 
Given this definition, the normalization of the conditional probability 
$\mu(\hat E_k|x,n)$ and its non-negativity can be stated as follows.
\be
\label{cond_sum_prob}
\begin{array}{l}
(\hat E_1,\cdots,\hat E_M)\in\Omega_n(x)\Rightarrow 
\vspace{1mm}
\\
\left\{
\begin{array}{l}
\sum_{k=1}^M \mu(\hat E_k|x,n)=1  
\vspace{1mm}
\\
\mu(\hat E_k|x,n)\ge 0 \;\;\;\; k\in\{1,\dots,M\}
\end{array}
\right.  .
\end{array}
\ee
This property, the normalization of $P_c(n|x,{\cal M},\tau)$ and its non-negativity
guarantee that the 
probability distribution $P(\hat E_k|x,{\cal M},\tau)$ defined by Eq. (\ref{cond_prob2}) is 
normalized and non-negative.

Since the family of sets $\Omega_n(x)$ is countable, we can replace these sets with
open sets by removing zero-measure boundaries. The resulting family is identical
to the original one up to a negligible zero-measure set of measurements.
Thus, we can just assume that the sets $\Omega_n(x)$ are open without loss of generality.
Furthermore, we can assume that they are connected. Indeed, if the sets are not connected,
we first can write them as union of connected sets, 
\be
\Omega_{n}(x)=\cup_k \Omega_{n,k}(x)
\ee
and replace the probability distribution $P_c(n|x,{\cal M},\tau)$ with
\be
P_c(n,k|x,{\cal M},\tau)\equiv P_c(n|x,{\cal M},\tau)\,\delta[{\cal M}\in \Omega_{n,k}(x)],
\ee
where $\delta[\text{true}]=1$ and $\delta[\text{false}]=0$. Then, we can rename
the pair $(n,k)$ by using only one discrete index $n$. In this way, we obtain a new
model and new corresponding sets $\Omega_n(x)$ that are open and connected.
Thus, we can assume that the sets $\Omega_n(x)$ satisfy the following. 
\vspace{2mm}
\newline
\begin{property}
\label{prop2}
The sets $\Omega_n(x)$ are open and connected for every $n$ and
every $x$.
\end{property}

A measurement $(\hat E_1 + \hat E_2,\dots,\hat E_M)\equiv {\cal M}_c$
can be implemented as the coarse graining of the measurement 
$(\hat E_1,\hat E_2,\cdots,\hat E_M) ={\cal M}$. Thus, we have the inference
\be\label{inference_coarse_gr}
{\cal M}\in\Omega_n(x)\Rightarrow {\cal M}_c \in  \Omega_n(x).
\ee 
Indeed, if ${\cal M}\in\Omega_n(x)$, then 
there is a context $\tau$ such that $P_c(n|x,{\cal M}_c,\tau)\ne 0$.
In general, the opposite inference is not true, that is,
\be
{\cal M}_c\in\Omega_n(x)\not\Rightarrow {\cal M} \in  \Omega_n(x).
\ee 
Indeed, the measurement ${\cal M}_c$ could be implemented without involving a 
coarse graining of ${\cal M}$. Thus, the two measurements could be associated with 
different values of the context index $n$.

For the following discussions, it is useful to define the operators ${\cal P}_k$.
\begin{definition}
Let $S$ be a set of ordered $m$-tuples.  ${\cal P}_k S$ with $k\in\{1,\dots,m\}$ 
is a set such that an element $p$ is in ${\cal P}_k S$ if and only
if there is an $M$-tuple $b$ in $S$ whose $k$-th component is equal to $p$.
\end{definition}
Thus, the operator ${\cal P}_k$ is a kind of Cartesian projector.
Similarly, let us define the operators ${\cal P}_{kl}$.
\begin{definition}
Let $S$ be a set of ordered $m$-tuples.  ${\cal P}_{kl} S$ with $k,l\in\{1,\dots,m\}$ 
and $k\ne l$ 
is a set of pairs such that an element $(p_1,p_2)$ is in ${\cal P}_{k,l} S$ if and only
if there is an $M$-tuple $b$ in $S$ whose $k$-th and $l$-th components are equal to 
$p_1$ and $p_2$, respectively.
\end{definition}
By inference~(\ref{inference_coarse_gr}), we have that property~(\ref{cond_sum_prob})
is equivalent to the following ones
\be\label{cond1}
\hat E\in {\cal P}_k \Omega_n(x)\Rightarrow
\mu(\hat E|x,n)\ge 0,
\ee
\be\label{cond2}
\begin{array}{c}
(\hat E_1,\hat E_2)\in {\cal P}_{kl} \Omega_n(x)\Rightarrow 
\vspace{1mm}
\\
\mu(\hat E_1|x,n)+\mu(\hat E_2|x,n)=\mu(\hat E_1+\hat E_2|x,n),  
\end{array}
\ee
\be\label{cond3}
\mu(\hat{\mathbb{1}})=1.
\ee

\section{Generalized Gleason theorem}
\label{sec_gen_gleason}

Before introducing the generalized Gleason theorem, let us briefly review the original 
theorem~\cite{gleason}. In the axiomatic formulation of quantum mechanics, each outcome is labeled by 
a projector $\hat E$ and the probability of $\hat E$, say $\mu(\hat E)$, is given by 
the Born rule
\be\label{quant_prob}
\mu(\hat E)=\text{tr}(\hat E\hat\rho),
\ee
where $\hat\rho$ is the density operator representing the state of the quantum 
system. This measure satisfies the two properties
\bey
\label{pos}
\mu(\hat E)\ge0,  \\
\label{sum}
\sum_{i=1}^M\mu(\hat E_i)=1,
\eey
where $\{\hat E_1,\cdots,\hat E_M\}$ is any complete set of commuting projectors
(so that $\sum_{k=1}^M \hat E_k=\hat{\mathbb{1}}$). 
Eq.~(\ref{sum}) is equivalent to the following conditions,
\bey
\label{identi}
&\mu(\hat{\mathbb{1}})=1,&  \\
\label{partial_sum}
&\begin{array}{c} 
\text{for every pair $\{\hat E_1,\hat E_2\}$ of commuting projectors}
\vspace{1.5mm}   \\ 
\Rightarrow
\mu(\hat E_1)+\mu(\hat E_2)=\mu(\hat E_1+\hat E_2).
\end{array} &  
\eey
Provided that the Hilbert space dimension is greater than $2$, Gleason's theorem 
states that any measure with properties (\ref{pos}) and (\ref{sum}) [or, equivalently,
properties (\ref{pos},\ref{identi},\ref{partial_sum})] has the form 
(\ref{quant_prob}), where $\hat\rho$ is a non-negative operator with trace one. 
This result provides a way for reducing the 
axiomatic basis of quantum mechanics. Indeed, it shows that the Born rule can be 
inferred by the assumption that every outcome is associated with a projector and every 
complete set of commuting projectors represents a complete set of measurement outcomes. 
\begin{theorem} {\rm (Gleason's theorem)}
For a Hilbert space of dimension greater than $2$, a measure $\mu(\hat E)$ that satisfies 
properties (\ref{pos}) and (\ref{sum}) has the form $\mu(\hat E)=\text{tr}(\hat E \hat\rho)$, 
where $\hat\rho$ is a Hermitian non-negative definite matrix with trace equal to $1$.
Equivalently, the same conclusion holds if properties~(\ref{pos},\ref{identi},\ref{partial_sum})
are satisfied.
\end{theorem}

Now, we present a generalization that has weaker hypotheses than Gleason's theorem 
and the almost identical conclusion. It requires that the Gleason hypotheses hold 
locally in some open subset of $\Omega$. We only introduce the additional hypothesis 
that $\mu(\hat E)$ is a generalized function (mathematical distribution)~\cite{friedman}
for which the derivatives are 
well-defined in the domain of $\mu$. Indeed, this can be considered 
the only case that is physically relevant. 
It is worth stressing that we are not assuming the stronger hypothesis of differentiability. 
Indeed, a piecewise differentiable function with discontinuities along some zero-measure 
subset is an example of mathematical distribution. More generally, integrable functions 
on compact sets are physically relevant examples of distributions. A distribution is formally 
defined as a functional from a set of test functions to $\mathbb{R}$.  With an abuse of notation, we 
will represent distributions as conventional functions. Our hypothesis on the function $\mu$
is complementary to the non-negativity hypothesis used by Gleason~\cite{gleason}, which is not required 
by the generalized Gleason theorem. The latter property, in the original theorem, rules 
out highly discontinuous unbounded functions satisfying the addition rule in inference~(\ref{partial_sum}).

\begin{theorem} {\rm(Generalization of Gleason's theorem I)}
\label{gen_gl_th0}
Let $O$ be a connected open set of 
complete $M$-tuples of commuting projectors, say $(\hat E_1,\dots,\hat E_M)$,
with $M>2$.
Let $\mu(\hat E)$ be a generalized function whose derivatives are
well-defined in $\cup_i{\cal P}_i O$. If the equality 
\be
\sum_i\mu(\hat E_i)=1
\ee
is satisfied for every $M$-tuple in $O$, then there is an Hermitian operator
$\hat\eta$ and real numbers $K_1,\dots,K_M$ such that
\be
\hat E\in{\cal P}_i{O} \Rightarrow
\mu(\hat E)=\text{tr}(\hat\eta\hat E)+K_i.
\ee
If the intersection of ${\cal P}_i{O}$ and ${\cal P}_j{O}$ is not empty,
then $K_i=K_j$.
\end{theorem}

The theorem can be equivalently stated as follows.
\begin{theorem} {\rm(Generalization of Gleason's theorem II)}
\label{gen_gl_th}
Let $O$ be a connected open set 
of incomplete pairs of commuting projectors.
Let $\mu(\hat E)$ be a distribution whose derivatives are
well-defined in ${\cal P}_1 O\cup {\cal P}_2 O$.
If the property
\be
\mu(\hat E_1)+\mu(\hat E_2)=\mu(\hat E_1+\hat E_2),
\ee
is satisfied for every pair  $(\hat E_1,\hat E_2) \in O$,
then there is a Hermitian operator $\hat\eta$ such that 
\be\label{gen_result}
\hat E\in{\cal P}_i{O} \Rightarrow
\mu(\hat E)=\text{tr}(\hat\eta\hat E)+K_i\equiv tr(\hat\eta_i\hat E),
\ee
where $\hat\eta_i=\hat\eta+(K_i/r_i)\hat{\mathbb{1}}$, $r_i$ being
the rank of $\hat E\in {\cal P}_i O$.
If the intersection of ${\cal P}_1{O}$ and ${\cal P}_2{O}$ is not empty,
then $K_1=K_2$ and $r_1=r_2$, so that
\be
\hat E\in{\cal P}_i{O} \Rightarrow
\mu(\hat E)=tr(\hat\eta\hat E)
\ee
for some Hermitian operator $\hat\eta$.
\end{theorem}
It is worth to remark that the operator $\hat\eta$ in Eq.~(\ref{gen_result})
does not depend on the index $i$.
Note that, if the additional hypothesis $\mu(\hat E_i)\ge0$ is added,
the density operator $\hat\eta_i=\hat\eta+(K_i/r_i) \hat{\mathbb{1}}$ is not 
necessarily non-negative
defined. Indeed, the function $\text{tr}(\hat\eta_i\hat E)$
must be positive only in a subset of projectors.
The two theorems are equivalent. Indeed, it is clear that Theorem~\ref{gen_gl_th}
implies Theorem~\ref{gen_gl_th0}. The other inference comes by taking
$\hat E_3\equiv \hat{\mathbb{1}}-\hat E_1-\hat E_2$ and 
$\mu(\hat E_3)\equiv 1-\mu(\hat E_1)-\mu(\hat E_2)$ for $(\hat E_1,\hat E_2)\in O$.

To prove Theorem~\ref{gen_gl_th}, we first consider projections
onto one-dimensional spaces. Thus, we assume that 
$\hat E_i\equiv \vec\phi_i\vec\phi_i^\dagger$ for $i=1,2$, where 
$\vec\phi_1$ and $\vec\phi_2$ are two unit orthogonal column vectors. 
We denote by $\phi_{i;k}$ the components of $\vec\phi_i$ and
define the function 
$$
f(\vec\phi_i)\equiv\mu(\vec\phi_i{\vec\phi_i}^{\hspace{0.4mm}\dagger}),
$$
which is called by Gleason {\it frame function}. Under the restriction
of rank-$1$ projectors, Theorem~\ref{gen_gl_th} takes the form of the following. \newline
\begin{lemma}
\label{lemma_ff}
Given a Hilbert space of dimension larger than $2$, let $\cal O$ be an open 
set of ordered pairs of orthogonal vectors. If, for any pair $(\vec\phi_1,\vec\phi_2)
\in\cal O$, the frame function satisfies the properties
\bey
f(\vec\phi_i)\ge0, \\
\label{sum_frames}
f(\vec\phi_1)+f(\vec\phi_2)=\mu(\vec\phi_1{\vec\phi_1}^{\hspace{0.4mm}\dagger}+
\vec\phi_2{\vec\phi_2}^{\hspace{0.4mm}\dagger}),
\eey
then the third-order derivatives of $f(\vec \phi)$ with respect to $\vec\phi$ are
equal to zero in ${\cal P}_1{\cal O}\cup{\cal P}_2{\cal O}$. In particular, if 
$\cal O$ is connected, then
there is a Hermitian operator $\hat\eta$ and two constants $K_1$ and $K_2$
such that
\be
\vec\phi\in{\cal P}_i{\cal O}\Rightarrow f(\vec\phi)=tr(\hat\eta  \vec\phi\vec\phi^\dagger) +K_i
\ee
for $i=1,2$.
\end{lemma}
Equation (\ref{sum_frames}) states that the sum $f(\vec\phi_1)+f(\vec\phi_2)$ depends
only on the subspace spanned by the vectors $\vec\phi_1$ and $\vec\phi_2$. Thus, if
the pair $(\vec\chi_1,\vec\chi_2)$ is in $\cal O$ and the vectors $\vec\chi_i$ are
linear combinations of $\vec\phi_i$, then $f(\vec\phi_1)+f(\vec\phi_2)=
f(\vec\chi_1)+f(\vec\chi_2)$.

Note that $f(\vec\phi)$ is defined on the unit sphere. As we will see, it is useful 
to expand the domain of $f$ to the whole vector space and introduce the radial constraint
\be\label{radial_constr}
\vec\phi\cdot\frac{\partial f}{\partial\vec\phi}+
\vec\phi^*\cdot\frac{\partial f}{\partial\vec\phi^*}=2 f.
\ee
It is always possible to expand the domain and satisfy this constraint with a suitable
choice of the radial behavior of $f$. Indeed, given a function $f(\vec\phi)$ on the unit 
sphere, the function $f\left(\frac{\vec\phi}{|\vec\phi|}\right)|\vec\phi|^2$ on the vector 
space is equal to $f(\vec\phi)$ on the unit sphere and satisfies Eq.~(\ref{radial_constr}).
\newline
{\bf Proof of Lemma \ref{lemma_ff}}.
The main task is to prove that
\be\label{complex_3der}
\begin{array}{c}
\frac{\partial^3 f(\vec\phi)}{\partial\phi_i\partial\phi_j\partial\phi_k}=0, 
\vspace{1mm}\\
\frac{\partial^3 f(\vec\phi)}{\partial\phi_i\partial\phi_j\partial\phi_k^*}=0
\end{array}
\ee
and their complex conjugations. For this purpose, it is sufficient to 
prove the real version of these equalities for a real three-dimensional space.

For any tern $\{i_1,i_2,i_3\}$ of integers such that $i_1\ne i_2, i_3$ and 
$i_2\ne i_3$, 
we write the components $\phi_{1;i_k}$ and $\phi_{2;i_k}$ in the form
\be
\phi_{1;i_k}\equiv v_k e^{i\varphi_k},
\phi_{2;i_k}\equiv w_k e^{i\varphi_k} \;\;\; k\in\{1,2,3\},
\ee
where $v_k$ and $w_k$ are components of two orthogonal three-dimensional 
real vectors, $\vec v$ and $\vec w$ respectively.
The task is reduced to prove that
\be\label{third_ders}
\frac{\partial^3 f(\vec v)}{\partial v_i\partial v_j\partial v_k}=0, 
\text{  for any } i,j,k\in\{1,2,3\}.
\ee
The generator of a three-dimensional rotation of $\vec v$ and $\vec w$ is
\be
{\cal R}(\vec a)=\vec a\cdot\left(\vec v\wedge\frac{\partial}{\partial\vec v}+
\vec w\wedge\frac{\partial}{\partial\vec w}\right)
\ee
where $\vec a$ gives the rotation axis. For a rotation in the plane
spanned by the orthogonal vectors $\vec v$ and $\vec w$, we have that
$\vec a=\vec v\wedge\vec w$. From Eq.~(\ref{sum_frames}), we have that
${\cal R}(\vec v\wedge\vec w)[f(\vec v)+f(\vec w)]=0$, that is,
\be\label{const_rot}
\left(\vec w\cdot\frac{\partial}{\partial\vec v}-\vec v\cdot\frac{\partial}
{\partial\vec w} \right)\left[f(\vec v)+f(\vec w)\right]=0
\ee
for every pair of orthogonal vectors $\vec v$ and $\vec w$ in the domain
of definition of $f$.
The generators of the rotations around $\vec w$ and $\vec v$ are 
$(\vec v\wedge\vec w)\cdot\frac{\partial}{\partial\vec v}$ and
$(\vec v\wedge\vec w)\cdot\frac{\partial}{\partial\vec w}$, respectively.
Applying these operators to both sides of Eq. (\ref{const_rot}),
we obtain the two equations 
\bey\label{op2_1}
\sum_{ij}(\vec v\wedge\vec w)_i w_j\frac{\partial^2 f(\vec v)}{\partial v_i\partial v_j}=
(\vec v\wedge\vec w)\cdot\frac{\partial}{\partial\vec w} f(\vec w), \\
\label{op2_2}
\sum_{ij}(\vec v\wedge\vec w)_i v_j\frac{\partial^2 f(\vec w)}{\partial w_i\partial w_j}=
(\vec v\wedge\vec w)\cdot\frac{\partial}{\partial\vec v} f(\vec v).
\eey
Then, we apply again the operator $\vec w\cdot\frac{\partial}{\partial\vec v}-
\vec v\cdot\frac{\partial}{\partial\vec w}$ to both sides of Eq. (\ref{op2_1})
and obtain 
\be\begin{array}{c}
\sum_{ij}(\vec v\wedge\vec w)_i \left(v_j\frac{\partial^2 f(\vec v)}{\partial v_i\partial v_j}-
 w_j \sum_k w_k\frac{\partial^3 f(\vec v)}
{\partial v_i\partial v_j\partial v_k}\right) =    \vspace{1mm} \\
\sum_{ij}(\vec v\wedge\vec w)_i v_j\cdot\frac{\partial^2 f(\vec w)}{\partial w_i\partial w_j}.
\end{array}
\ee
Thus, the left-hand side of this equation is equal to the right-hand side
of Eq. (\ref{op2_2}), that is,
\be
\label{third_der0}
\begin{array}{c}
\sum_{ijk}(\vec v\wedge\vec w)_i w_j w_k\frac{\partial^3 f(\vec v)}
{\partial v_i\partial v_j\partial v_k}=  \vspace{1mm} \\
(\vec v\wedge\vec w)\cdot\frac{\partial}{\partial\vec v}
\left[\vec v\cdot\frac{\partial f(\vec v)}{\partial\vec v}-2 f(\vec v) \right].
\end{array}
\ee
From Eq. (\ref{radial_constr}) we have that
\be\label{radial_real}
\vec v\cdot\frac{\partial f(\vec v)}{\partial\vec v}-2 f(\vec v)=0.
\ee
Thus, Eqs. (\ref{third_der0}) and (\ref{radial_real}) imply that
\be
\label{third_der1}
\sum_{ijk} u_i w_j w_k\frac{\partial^3 f(\vec v)}
{\partial v_i\partial v_j\partial v_k}=0
\ee
for every tern  $\{\vec u,\vec v,\vec w\}$ of orthogonal vectors. This implies 
that
\be
\label{third_der2}
\sum_{ijk} w_i w_j w_k\frac{\partial^3 f(\vec v)}
{\partial v_i\partial v_j\partial v_k}=0
\ee
for every pair $\{\vec v,\vec w\}$ of orthogonal vectors. 
Indeed, this equation can be derived from Eq.~(\ref{third_der1}) 
by considering the two pairs of orthogonal vectors 
$(\vec u,\vec w)=(\vec u'\mp\vec w',\vec u'\pm\vec w')$, 
where $\vec u'$ and $\vec w'$ are vectors orthogonal
to $\vec v$ and with $|\vec u'|=|\vec w'|$. These two cases
and the equations
$$
\sum_{ijk} u_i' w_j' w_k'\frac{\partial^3 f(\vec v)}
{\partial v_i\partial v_j\partial v_k}=
\sum_{ijk} u_i' u_j' w_k'\frac{\partial^3 f(\vec v)}
{\partial v_i\partial v_j\partial v_k}=0
$$
give the two equations $(w_i' w_j' w_k'\pm  u_i' u_j' u_k' )
\frac{\partial^3 f}{\partial v_i\partial v_j\partial v_k}= 0$, which imply
Eq.~(\ref{third_der2}).
Thus, every third-order derivative in the subspace orthogonal to $\vec v$ 
is equal to zero. Furthermore, from Eq.~(\ref{radial_real}) we have that
\be
\vec v\cdot\frac{\partial^3 f(\vec v)}{\partial\vec v\partial v_i\partial v_j}=0
\ee
for every $i,j\in\{1,2,3\}$.
This implies that every third-order derivative is zero and, thus, Eq.~(\ref{third_ders}) 
is satisfied. Identical equations hold for $f(\vec w)$.
Since Eq. (\ref{third_ders}) holds for any real three-dimensional subspace of 
a complex Hilbert space, also Eqs.~(\ref{complex_3der}) and their complex conjugations
hold. Thus, the function $f(\vec\phi_i)$ must be quadratic in $\vec\phi_i$. 
Since the frame function $f(\vec\phi_i)$ is equal to $\mu(\vec\phi_i\vec\phi_i^\dagger)$,
the linear terms and the terms in $\phi_{i;k}\phi_{i;l}$ and $\phi_{i;k}^*\phi_{i;l}^*$
are equal to zero. In particular, if $\cal O$ is connected, then
\be
f(\vec\phi)= tr(\hat\eta  \vec\phi\vec\phi^\dagger) +K_i
\ee
for $\vec\phi\in{\cal P}_i{\cal O}$ and $i\in\{1,2\}$.
The lemma is proved. $\square$

{\bf Proof of Theorem \ref{gen_gl_th}}. We can decompose the two projectors
$\hat E_1$ and $\hat E_2$ into rank-$1$ commuting projectors, say $F_i^{(k)}$,
\be
\hat E_1=\sum_{k=1}^{r_1} \hat F_1^{(k)},  \;\;  \hat E_2=\sum_{k=1}^{r_2} \hat F_2^{(k)},
\ee 
where $r_i$ is the rank of $\hat E_i$. Since the pair $(\hat E_1,\hat E_2)$
is not complete, we have that $r_1+r_2<N$, where $N$ is the dimension of
the Hilbert space. Let us denote by $\vec \phi_i^k$ the vectors such
that $\hat F_i^{(k)}=\vec \phi_i^k (\vec\phi_i^k)^\dagger$.
By Lemma~(\ref{lemma_ff}), we have that the function $\mu$ must be
linear in $\hat F_i^{(k)}$ in each connected set 
for any decomposition of $\hat E_1$ and $\hat E_2$.
For example, keeping $\hat F_1^{(2)},\cdots,\hat F_1^{(r_1)}$ and
$\hat F_2^{(2)},\cdots,\hat F_2^{(r_2)}$ constant. The orthogonal complement,
say ${\cal H}_\perp$, of the subspace spanned by 
$\vec\phi_1^2,\dots,\vec\phi_1^{r_1}$ and
$\vec\phi_2^2,\dots,\vec\phi_2^{r_2}$ is a vector subspace of dimension
equal to $N-r_1-r_2+2>2$. We denote by $\Pi$ the set of pairs of orthogonal
vectors in ${\cal H}_\perp$. The set of pairs $(\vec\phi_1^1,\vec\phi_2^1)$
such that $(\hat E_1,\hat E_2)\in O$ is an open set of $\Pi$. Thus, by Lemma~\ref{lemma_ff},
the third-order derivatives of $\mu$ with respect to $\vec\phi_1^1$ and
$\vec\phi_2^1$ are equal to zero.
This is true for every decomposition of $\hat E_i$ into
rank-$1$ projectors.
This implies that $\mu$ is linear in $\hat E_1$ and $\hat E_2$ and has
the form 
\be
\mu(\hat E_i)=\text{tr}(\hat\eta\hat E_i)+K_i
\ee
for any $\hat E_i\in{\cal P}_i O$. If the intersection of ${\cal P}_1 O$ and
${\cal P}_2 O$ is not empty, it is trivial that $K_1=K_2$ and $r_1=r_2$. $\square$

Gleason's theorem is a trivial consequence of theorem~\ref{gen_gl_th}.
Indeed, if the set $O$ in the theorem statement is the whole set
of pairs of commuting projectors with fixed rank $r$, then the coefficients 
$K_i$ in 
Eq.~(\ref{gen_result}) are independent of $i$, $K_i=K$. Let us define 
$\hat\rho\equiv\hat\eta +K r^{-1} \hat{\mathbb{1}}$, Eq.~(\ref{gen_result}) gives
\be
\mu(\hat E)=tr (\hat\rho \hat E).
\ee
The non-negativity of $\mu(\hat E)$ and the equality
$\mu(\hat{\mathbb{1}})=1$ imply that $\hat\rho$ is positive 
semidefinite and with trace equal to $1$.

\section{Functional form of the outcome probability in a HV theory}
\label{sec_prob_meas}


The generalization of the Gleason theorem has an obvious consequence for the 
functional form of the conditional probability $\mu(\hat E|x,n)$ defined
in Sec. \ref{sec_context}. Let us remind that this function satisfies the 
three conditions
\bey
\nonumber
&\hat E\in{\cal P}_k\Omega_n(x)\Rightarrow \mu(\hat E|x,n)\ge0,&  \\
\nonumber
&\begin{array}{c}
(\hat E_1,\hat E_2)\in{\cal P}_{kl}\Omega_n(x)\Rightarrow \vspace{1mm} \\
\mu(\hat E_1|x,n)+\mu(\hat E_2|x,n)=\mu(\hat E_1+\hat E_2|x,n),
\end{array}&   \\
\nonumber
&\mu(\hat{\mathbb{1}}|x,n)=1.&  
\eey
stated in Eqs. (\ref{cond1},\ref{cond2},\ref{cond3}). In particular, the
second hypothesis is identical to that used for the generalized Gleason
theorem. Indeed, according to property~\ref{prop2}, the set ${\cal P}_{kl}\Omega_n(x)$
 is a connected open set apart from a negligible boundary. Thus, the generalized 
Gleason theorem implies that 
\be\label{cond_prob_lin}
\hat E\in {\cal P}_k \bar \Omega_n(x)  \Rightarrow
\mu(\hat E|x,n)=\text{tr}\left[\hat\eta(x,n)\hat E\right]+K(x,n),
\ee
where $\bar\Omega_n(x)$ is the subset of $\Omega_n(x)$ containing
measurements with three or more outcomes.
Inferences~(\ref{cond_sum_prob}) also imply that
\be\begin{array}{c}
\{\hat E_1,\cdots,\hat E_M\}\in\bar\Omega_n(x)\Rightarrow  \vspace{1mm} \\
\sum_{k=1}^M \left\{\text{tr}[\hat E_k\hat\eta(x,n)]+K(x,n)\right\}=1,
\end{array}
\ee
\be\begin{array}{c}
\{\hat E_1,\cdots,\hat E_M\}\in\bar\Omega_n(x)\Rightarrow  \vspace{1mm} \\
\text{tr}[\hat E_k\hat\eta(x,n)]+K(x,n)\ge 0.
\end{array}
\ee

Note that an outcome-deterministic theory is compatible with the equations 
that we have derived from the generalized Gleason theorem. Indeed, the 
ontological theory is outcome-deterministic if, for example, 
$\hat\eta(x,n)=0$ and $K(x,n)$ is identically equal to $0$ or $1$ where
$P_c(n|x,{\cal M},\tau)\rho(x|\psi,\eta)$ is different from zero.

In fact, we have not proved that there 
is an ontological model such that the context information is finite,
we have only shown that, if it exists, then it must 
have some general structure. However, an approximate classical protocol 
simulating entanglement and quantum channels, reported in 
Ref.~\cite{montina2}, would suggest that such a model exists.
In this section, we have assumed that the measurement is performed all at once. 
This justifies why the context for the event $\hat E_k$, in general, depends on 
the whole set {\cal M} of projectors. 
Given multiple commuting measurements, causality imposes some further
constraints. Suppose that two-outcome measurements are performed consecutively 
with outcomes $\{\hat E_1,\hat{\mathbb{1}}-\hat E_1\}$, 
$\{\hat E_2,\hat{\mathbb{1}}-\hat E_2\}$,..., $\{\hat E_M,\hat{\mathbb{1}}-\hat E_M\}$,
where $\hat E_k$ are commuting projectors. Under the hypothesis of causality, the 
outcome of a measurement cannot be influenced by future measurements. 
Thus, we can rearrange the context index $n$ as an $M$-tuple of indices $(n_1,\dots,n_M)\equiv\vec n$
so that the conditional probability $\mu$ of $\hat E_k$ given $\vec n$ depends only
on the first $m$ indices, that is,
\be
\begin{array}{l}
\mu(\hat E_1|x,\vec n)=\mu(\hat E_1|x,n_1)  \\
\dots   \\
\mu(\hat E_M|x,\vec n)=\mu(\hat E_M|x,n_1,\dots,n_M),
\end{array}
\ee
and the conditional probability of the first $k$ indices $n_1,\dots,n_k$ depends only
on the first $k$ projectors, that is,
\be
\begin{array}{c}
P_c(n_1,\dots,n_k|x,{\cal M},\tau)=  \\
P_c(n_1,\dots,n_k|x,\hat E_1,\dots,\hat E_k,\tau),  \;\;\; k\in\{1,\dots,M\}.
\end{array}
\ee

We conclude this section by discussing a relation between the hypothesis of
finiteness of the contextual information and a long-standing debate about the 
nature of the quantum state, which reached its apex with the 
Pusey-Barrett-Rudolph theorem~\cite{pbr}. In the framework of onthological 
theories, we can distinguish two possible cases. In one case, the quantum
state is part of the classical description, so that
the quantum state can be inferred by knowing the ontic state $x$. More
precisely, two distributions $\rho(x|\psi,\eta)$ and $\rho(x|\psi',\eta')$  
with $\psi\ne\psi'$ are not overlapped. In the other case, the inference
of the quantum state from the ontic state is not generally possible.
In Ref.~\cite{pbr}, it was proved that the second case takes to a
contradiction under a hypothesis of separability. Namely, the PBR 
hypothesis states, shortly speaking, that two
spatially separate systems prepared in two quantum states (so that
the overall quantum state is the product of the states) is associated with
statistically independent classical variables. If the quantum state is taken
as part of the classical description, then there are scenarios involving
multiple measurements such that the information about the context is
infinite. Thus, our hypothesis of finiteness of information has to
lead to a break of the PBR separability condition under some scenario
involving distinct systems and multiple measurements.

\section{Conclusion and perspectives}

In this paper, we have presented a generalization of the Gleason theorem and 
illustrated its application by deriving some general properties of a special 
class of HV theories. Apart from their relevance in quantum foundations,
these theories are important also in quantum communication complexity as
classical simulation protocols of quantum channels~\cite{toner,massar}.
Assuming that the amount of relevant information about 
the measurement context is finite, we have proved that the probability of an 
event for a single measurement with more than two outcomes must be linear in 
the projector associated with the event, given the information about 
the context. Further properties can be deduced considering multiple commuting 
measurements under the assumption of causality. This generalization of the 
Gleason theorem can suggest some clues for finding a classical model that 
replaces the quantum communication of $n$ qubits with a finite amount of
classical communication. At the present, this model is missing, apart from the 
Toner-Bacon model for single qubit~\cite{toner} and a two-way communication model 
reported in Ref.~\cite{massar}. We found the lower bound $2^n-1$ for 
the amount of classical one-way communication required by an exact 
simulation~\cite{montina3} of $n$ qubits. We have also discussed a possible relation 
of this work with the long-lasting debate on the nature of the quantum
state~(see Ref.~\cite{pbr} and references in there).

We conclude by suggesting some other possible extensions of this work. The proof 
of the generalized Gleason theorem requires that the measure $\mu$  is a 
generalized function~\cite{friedman}, for which the derivatives are well-defined. 
This condition replaces the non-negativity condition used by Gleason, which has 
the same effect of ruling out highly discontinuous functions. Although our 
hypothesis is physically reasonable, it make the generalized Gleason theorem 
partially complementary to the original theorem. It would be interesting to find 
an alternative proof that requires only the non-negativity hypothesis and possibly 
uses an even weaker hypothesis on the set $O$ (see theorem~\ref{gen_gl_th}). 
It is worth noting that the concept of contextuality also 
applies to the preparation procedure~\cite{spekkens}, thus
we could wonder if also this kind of contextuality can be summarized by a finite 
amount of information. In such a case, we would find that, given this partial 
information on the preparation context, the probability distribution of the 
ontic state $x$ should be quadratic in the quantum state, like  a quasi-probability 
distribution (such as Glauber-Sudarshan $P$ distribution).
In other terms, the probability distribution $\rho(x|\psi,\eta)$ would be somehow piecewise 
quadratic. Indeed, in Ref.~\cite{montina4} we proved that $\rho(x|\psi,\eta)$ cannot be 
quadratic on the whole Hilbert space, which is equivalent to say that a HV theory is contextual 
for a state preparation procedure, as remarked in Ref.~\cite{spekkens2}.
  
\section{Acknowledgments}
We thank Xavier Coiteux-Roy for a careful reading of the manuscript and helpful
comments. This work was supported by Swiss National Science Foundation (SNF).

\end{document}